\documentclass[aps,prd,nofootinbib,showpacs,twocolumn,floatfix,superscriptaddress]{revtex4}
\usepackage{amssymb,amsfonts}
\usepackage{bm} 
\usepackage{graphicx}

\newcommand{\der}[2]{\frac{\partial #1}{\partial #2}}

\newcommand{\w}[1]{\bm{#1}}
\newcommand{\be}{\begin{equation}}
\newcommand{\ee}{\end{equation}}
\newcommand{\bea}{\begin{eqnarray}}
\newcommand{\eea}{\end{eqnarray}}
\newcommand{\Hor}{{\mathcal H}}

\newcommand{\Lie}[1]{\bm{\mathcal L}_{\w{#1}}\,}

\newcommand{\tD}{\tilde D}

\begin{document}

\title{Trapping Horizons as inner boundary conditions for 
black hole spacetimes }

\newcommand*{\IAA}{Instituto de Astrof\'{\i}sica de Andaluc\'{\i}a, CSIC, Apartado Postal 3004, Granada 
        18080, Spain} 
\newcommand*{\MEU}{Laboratoire Univers et Th\'eories (LUTH), Observatoire de Paris, CNRS, 
Universit\'e Paris Diderot ; 5 place Jules Janssen, 92190 Meudon, France} 
\newcommand*{\VAL}{Departamento de Astronom\'\i a y Astrof\'\i sica, 
Universidad de Valencia, Valencia, Spain}

\author{Jos\'e Luis Jaramillo}
\email{jarama@iaa.es} 
\affiliation{\IAA}
\affiliation{\MEU}
\author{Eric Gourgoulhon}
\email{eric.gourgoulhon@obspm.fr}
\affiliation{\MEU}
\author{Isabel Cordero-Carri\'on}
\email{Isabel.Cordero@uv.es} 
\affiliation{\VAL}
\author{Jos\'e Mar\'\i a Ib\'a\~nez}
\email{Jose.M.Ibanez@uv.es} 
\affiliation{\VAL}

\date{18 september 2007}

\begin{abstract}
We present a set of inner boundary conditions for the numerical construction
of dynamical black hole space-times, when employing a 3+1 constrained 
evolution scheme and an excision technique. 
These inner boundary conditions are heuristically motivated by the
dynamical trapping horizon framework and are enforced in an 
elliptic subsystem of the full Einstein equation.
In the stationary limit they reduce to existing isolated horizon boundary conditions.
A characteristic analysis completes the  discussion of
inner boundary conditions for the radiative modes. 
\end{abstract}


\maketitle


\noindent\emph{General problem.}
The aim of this report is to discuss a set of inner boundary conditions (BC)
for dynamical evolutions of black hole spacetimes using an excision technique.
These BCs are derived in the context of 
the dynamical trapping horizon framework \cite{Haywa,AshteK,AshteK04,Booth05}.
In parallel with the recent black hole
numerical studies based on free evolution schemes,
which have led to the successful simulations of binary 
black hole coalescence through the merger phase (see e.g. \cite{BakerCP_al07} for
extensive references),
a 3+1 scheme for a fully-constrained evolution of Einstein equation has been
presented in Ref. \cite{BonazGGN04}. This approach 
maximizes the number of elliptic equations to be solved 
during the evolution, resulting in a coupled elliptic-hyperbolic
PDE system \cite{CordeIJG07}. Spectral methods \cite{GrandN07} are then employed
both to solve the elliptic subsystem and to handle the spatial part of the
relevant hyperbolic operators. 
We deal with the black hole singularity by means of the excision technique.
This raises the question about the appropriate choice
of inner BCs on the excised sphere, both for the elliptic and
the hyperbolic parts of the system. Regarding the hyperbolic equations, this 
inner boundary issue is intimally related
to the metric type of the world-tube hypersurface generated by the 
time evolution of the excision
sphere. As observed in Ref. \cite{ScheelPL_al06}, certain choices for the excision
surface render this excision hypersurface partially time-like, leading to 
ill-posedness if inconsistent BCs are supplied for the 
radiative modes.
A solution to this problem is suggested by the quasi-local approach to the evolution of black hole horizons,
embodied in the \emph{dynamical trapping horizon} framework 
(see review articles \cite{AshteK04,Booth05} and also Ref. \cite{BoothF07}).
This formalism motivates a natural
geometric choice for the excision surface. 
The basic underlying idea goes back to Eardley's work \cite{Eardl98} and consists in 
modeling the black hole 
horizons by $S^2\times \mathbb{R}$ world-tubes sliced by 
{\em apparent horizons},
that satisfy certain additional conditions guaranteeing the
physical growth of the horizon area (see below). 
On the one hand, apparent horizons at each given 3-slice of the time evolution
provide non-ambiguous geometric choices for the excision sphere 
that are guaranteed to lay inside the event horizon, and therefore are causally
disconnected from the rest of the spacetime.
On the other hand, dynamical trapping horizons are space-like
hypersurfaces suggesting that no conditions must be supplied
at the inner boundaries for the modes propagating in the bulk.
In sum, this proposal recasts Eardley's program \cite{Eardl98} in the
dynamical trapping horizon setting.
In the following we describe the fully-contrained scheme, 
then we present inner BCs for the elliptic part that
guarantee that the excised sphere generates a (dynamical) trapping horizon, and 
finally we show that the combination of a Dirac-like gauge \cite{BonazGGN04}
and dynamical trapping horizon inner BCs for the elliptic part
of the PDE system, 
actually imply that no BCs must be prescribed for the
hyperbolic fields at the inner excised sphere.

\noindent\emph{Fully-constrained evolution scheme.}
In the setting of the standard 3+1 decomposition of a spacetime $({\cal M}, \w{g})$
by spatial slices $(\Sigma_t)$, Ref. \cite{BonazGGN04} proposes a particular initial-boundary
problem for the spacetime evolution from an initial Cauchy slice. 
Let us denote by $\w{n}$ the unit timelike normal vector to $\Sigma_t$, the spatial
3-metric by $\w{\gamma}$, i.e. $\w{\gamma}=\w{g}+\w{n}\otimes \w{n}$, and define the
extrinsic curvature of $\Sigma_t$ as $\w{K}=-\frac{1}{2}\Lie{n}\w{\gamma}$. The evolution 
vector $\w{t}\equiv \partial_t$ is decomposed in terms of the lapse function $N$ and the shift 
vector $\w{\beta}$, as $\w{t}= N\w{n}+\w{\beta}$.
In addition, we introduce a fidutial flat metric $\w{f}$, satisfying $\Lie{t}\w{f}=\partial_t f_{ij}=0$.
Now we proceed by performing
a conformal decomposition of the 3+1 fields: $\w{\gamma} =\Psi^4\tilde{\w{\gamma}}$,
$\w{K} =\Psi^4\tilde{\w{A}}+\frac{1}{3}K\w{\gamma}$,
where $K=\gamma^{ij}K_{ij}$, the representative $\tilde{\w{\gamma}}$ of the conformal class of the 3-metric is chosen
to be unimodular, i.e. satisfies $\mathrm{det}(\tilde{\w{\gamma}})=\mathrm{det}(\w{f})$,
and the traceless part $\tilde{\w{A}}$ of $\w{K}$ is written as
$\tilde{A}^{ij} = \frac{1}{2N}
        \left( \tilde{D}^i\beta^j + \tilde{D}^j\beta^i 
         - \frac{2}{3}  \tilde{D}_k\beta^k \tilde{\gamma}^{ij}
         + \partial_t{\tilde\gamma}^{ij}  \right)$, 
$\tilde{\w{D}}$ being the Levi-Civita connection associated with $\tilde{\w{\gamma}}$.
In a second step, a coordinate choice must be adopted. Following
the prescriptions in \cite{BonazGGN04}, namely \emph{maximal
slicing} and \emph{Dirac gauge}, we set
\be
\label{e:gauges}
K=0 \ \ , \ \ {\cal D}_k\tilde{\gamma}^{ki}=0 \ \ ,
\ee
where ${\cal \w{D}}$ is the Levi-Civita connection associated with the flat metric $\w{f}$
(see Ref.~\cite{Gourg07a} for a discussion and relation to other coordinate choices).
Conditions (\ref{e:gauges}) fix the coordinates up to boundary terms.
The Dirac gauge condition will play a key role in the following, whereas 
maximal slicing can be relaxed to an 
arbitrary $K$ vanishing asymptotically near
spacelike infinity.
Inserting the conformal decomposition and gauges (\ref{e:gauges})
into
Einstein equation results in a coupled elliptic-hyperbolic system \cite{BonazGGN04}.
The elliptic part can be written as
\bea
\label{e:elliptic_part}
 \tD_k \tD^k \Psi - \frac{{}^3\!{\tilde R}}{8} \, \Psi &=&
 S_{\Psi}[\Psi, N, \w{\beta}, \tilde{\w{\gamma}}] \nonumber \\
    \displaystyle \tD_k \tD^k \beta^i +  \frac{1}{3} \tD^i \tD_k \beta^k 
    + {}^3\!{\tilde R}^i_{\ \, k} \beta^k & = & S^i_{\beta}[\Psi,
 N, \w{\beta}, \tilde{\w{\gamma}}]   \\
    \displaystyle \tD_k \tD^k N + 2 \tD_k\ln\Psi\, \tD^k N 
    & = & 
 S_{N}[N, \Psi, \w{\beta},\tilde{\w{\gamma}}]  \nonumber \ \ ,
\eea
where the first equation on $\Psi$ follows from the Hamiltonian constraint,
and the equation for the shift $\w{\beta}$ results from the simultaneous
imposition of the preservation of the Dirac gauge in time, i.e. 
$\partial_t ({\cal D}_k\tilde{\gamma}^{ki})$=0,
together with 
the momentum constraint. The Dirac gauge ensures the elliptic
character of this equation. Finally the third equation follows from 
$\partial_t{K}=0$.
$S_{\Psi}$, $S_{\beta}$ and $S_{N}$ represent non-linear sources
given in Ref.~\cite{BonazGGN04}.
Note the similarity with the extended conformal thin sandwich elliptic system 
\cite{XCTS} for the construction of initial data. In the present context, Eqs.~(\ref{e:elliptic_part})
are meant to be solved along the whole evolution, not only
on an initial slice.
Regarding the evolution part, we solve for the deviation $\w{h}$ of the conformal
metric from the flat fidutial one $\w{f}$, i.e. $\w{h} = \tilde{\w{\gamma}} - \w{f}$.
We choose a second-order form for the evolution equations, 
that
can be formally written as
\be
\label{e:hyperbolic_part}
\frac{\partial^2 h^{ij}}{\partial t^2} - \frac{N^2}{\Psi^4}
\tilde{\gamma}^{kl}{\cal D}_k {\cal D}_l
h^{ij}
- 2 {\cal L}_{\beta}\frac{\partial h^{ij}}{\partial t} + {\cal
  L}_{\beta}{\cal L}_{\beta}h^{ij} = S_{h}^{ij} 
\ee
where the nonlinear sources 
$S_{h}^{ij}[N, \Psi, \w{\beta}, \tilde{\w{\gamma}}]$ do not contain 
second derivatives of $\w{h}$. 
Eqs. (\ref{e:elliptic_part}) and (\ref{e:hyperbolic_part})
are solved in Ref. \cite{BonazGGN04} inside a spacetime region 
bounded by an outer timelike tube at {\em large}
spatial distances. 
We focus here only on the
inner BC problem. On a first stage, 
dynamical trapping horizon considerations will provide inner BCs for 
the conformal factor $\Psi$, the shift $\w{\beta}$ and the lapse $N$. 
In a second step we will analyse the hyperbolicity of the subsystem (\ref{e:hyperbolic_part})
and, most importantly in the present context, we will evaluate its characteristics
fields and speeds to assess if inner BCs must be provided at all
for $\w{h}$. 

As mentioned above, we do not discuss here 
the important outer BC problem. 
In this sense, a very interesting alternative has been recently presented by 
Moncrief et al. \cite{Moncr07}. They propose a (conformal) 3+1 constrained scheme,
which differs crucially from \cite{BonazGGN04} in one point:
the chosen slicing, involving {\em constant mean curvature} slices, extends up to future null 
infinity ${\cal I}^+$, a natural boundary for physical outgoing radiation
conditions. 
This strategy 
permits to bypass the boundary problem at the outer timelike border.
The feature of \cite{Moncr07} we highlight in the context of the present 
work 
is the shared adoption of an inner excision approach to the black hole singularity problem.  
An alternative geometric choice for the inner surface is proposed in 
\cite{Moncr07}, namely
the use of minimal surfaces. However, our proposal of rather employing apparent horizons instead, 
straightforwardly translates also into their scheme.

\noindent\emph{Inner BCs for the elliptic part: dynamical trapping horizons.}
Quasi-local approaches to black hole horizons aim at modeling the boundary of a black hole region 
as world-tubes of apparent horizons $({\cal S}_t)$.
At each point of a given spacelike closed surface ${\cal S}_t$
we can define (up to total rescaling) two null vectors
 $\w{\ell}$ and $\w{k}$, satisfying $\w{k}\cdot \w{\ell}=-1$ and 
spanning the plane normal to ${\cal S}_t$. 
Denoting by $\w{q}$ the metric on ${\cal S}_t$ induced by the ambient metric $\w{g}$ and by 
$\w{\epsilon}_S$ the associated area element, we can define 
the expansion $\theta^{(\w{v})}$ and shear $\w{\sigma}^{(\w{v})}$
along any  vector $\w{v}$ normal to ${\cal S}_t$ by
$\Lie{v} \w{\epsilon}_S=\theta^{(\w{v})}\w{\epsilon}_S$ and 
$2\w{\sigma}^{(\w{v})}= \Lie{\w{v}}\w{q}- \theta^{(\w{v})} \w{q}$. 
The surface ${\cal S}_t$ is \emph{trapped} \cite{Penro65} if light rays emitted from it
locally converge:  $\theta^{(\w{k})}\leq0$ and $\theta^{(\w{\ell})}\leq 0$. 
In the limiting case in which one of the expansions vanishes, ${\cal S}_t$ is called 
a \emph{marginally trapped surface} (MTS).
Since we will deal with asymptotically flat 3-slices, we can unambiguosly
define an {\it outgoing} null normal, say $\w{\ell}$, as the one pointing towards spacelike infinity.   
Then, condition $\theta^{(\w{\ell})}=0$ defines a marginally {\it outer} trapped surface (MOTS) \cite{Hawki73}. In contrast with MTSs, MOTSs 
impose nothing on $\theta^{(\w{k})}$.
Apparent horizons are outermost MOTSs. 
In this context, quasi-local dynamical trapping horizons ${\cal H}$ are $S^2\times\mathbb{R}$ 
hypersurfaces sliced by MOTSs $({\cal S}_t)$ and satisfying $\theta^{(\w{k})}<0$. Actually,
slices $({\cal S}_t)$ are indeed MTSs but,
motivated by inner BCs below,
we wish to stress 
the underlying MOTS structure.
Following Hayward \cite{Haywa},
${\cal H}$ is a {\it future outer trapping horizon} (FOTH) if, in addition, 
$\Lie{k}\theta^{(\w{\ell})}<0$ holds. This represents a stability locally outermost condition,
essentially stating that the interior of ${\cal H}$ is a trapped region. 
FOTHs can be either
null o spacelike hypersurfaces, the former representing stationary situations and the latter
dynamical ones. Alternatively, {\it dynamical horizons} (DH) introduced by Ashtekar and Krishnan 
\cite{AshteK} substitute the
condition on $\Lie{k}\theta^{(\w{\ell})}$ by the requirement of ${\cal H}$ to be spacelike, stationarity 
being represented by (null)isolated horizons (IH). Both in FOTHs and DHs, condition  $\theta^{(\w{k})}<0$
guarantees that the horizon area is never decreasing. In the dynamical context, 
FOTHs and DHs have been shown to be equivalent \cite{AnderMS05,BoothF07}.
In our 3+1 description, slices $({\cal S}_t)$ of ${\cal H}$ will always lay within a spatial surface $\Sigma_t$
of the chosen $3+1$ slicing.
Denoting by $\w{s}$ the unit spacelike normal vector to ${\cal S}_t$ laying in $\Sigma_t$ and pointing
towards spacelike infinity,
we can perform a 2+1 decomposition on the horizon. In particular,
the metric $\w{q}$ induced on ${\cal S}_t$ can be written as $\w{q}=\w{\gamma}-\w{s}\otimes\w{s}$
and the shift can be decomposed in its normal and tangential part as: $\w{\beta}=\beta^\perp \w{s}-
\w{V}$, with $\beta^\perp =\w{\beta}\cdot \w{s}$ and $\w{V}\cdot \w{s}=0$.

\noindent A most important result in this context is the {\em foliation uniqueness
theorem} by Ashtekar and Galloway \cite{AshteG05} stating that, for a given DH ${\cal H}$, 
there exists a unique foliation $({\cal S}_t)$ by MTS's.
Using this, we can define a canonical vector $\w{h}$ as the vector tangent 
to $\Hor$, normal to each ${\cal S}_t$ 
and that Lie-drags each MTS ${\cal S}_t$ of $\Hor$
into another one ${\cal S}_{t+\delta t}$. It constitutes a natural evolution vector on ${\cal H}$
and can be decomposed as $\w{h} = N\w{n} + b \w{s}$, where the normalization $N$
follows from requiring ${\cal S}_t \in \Sigma_t$ and is fixed up to a factor only depending on $t$.
Defining a parameter $C$ as 
(half) the square norm of $\w{h}$ with respect to $\w{g}$, i.e. $C:=\w{h}\cdot\w{h}/2=b^2-N^2$, 
it follows from the above-commented 
metric type of FOTH's that $C\geq 0$; 
strict inequality $b-N>0$ holds
in the DH situation and 
$b-N=0$ in the equilibrium (null) IH case; accordingly, we normalize the null 
vector $\w{\ell}$ as the limit of $\w{h}$ in the
stationary case: $\w{\ell}=N(\w{n}+\w{s})$ \cite{Cook,JaramGM04}. 

\noindent Our criteria for setting BCs for Eqs. (\ref{e:elliptic_part}) are: a) to enforce the excision
world-tube ${\cal H}$ to be sliced by MOTS, and b) to recover IH 
BCs \cite{Cook,JaramGM04,DainJK05,GourgJ06,JaramAL07}
at the equilibrium limit $C=0$. Motivated by this second point, but ultimately justified 
by the inner boundary analysis of Eqs. (\ref{e:hyperbolic_part}), we choose a coordinate
system adapted to ${\cal H}$ by demanding $\w{t}$ to be tangent to  ${\cal H}$. This implies
$\beta^\perp=b$, and we have
\be
\label{e:adapted_coord}
\w{h} = \w{t}+ \w{V} \ \ , \ \ \beta^\perp - N \geq 0 \ \ .
\ee
\noindent \emph{i) Geometric conditions for ${\cal H}$.} 
The first two BCs are provided by 1) the geometric definition
of ${\cal S}_t$ as a MOTS: $\theta^{(\w{\ell})}=0$, 
and 2) the Lie-dragging of MOTS into MOTS by $\w{h}$ inside ${\cal H}$ ({\it trapping horizon} condition): $\Lie{h}\theta^{(\w{\ell})}=0$.
The first one yields
\be
\label{e:BC_Psi}
4 \tilde{\w{s}} \cdot\tilde{\w{D}}\mathrm{ln} \Psi +
\tilde{\w{D}}\cdot \tilde{\w{s}} + \Psi^{-2} K(\tilde{\w{s}},\tilde{\w{s}}) -
\Psi^2 K = 0 \ \ ,
\ee
where tildes refer to the conformal metric $\tilde{\w{\gamma}}$;
in particular, $\tilde{\w{s}}= \Psi^2 \w{s}$. The second geometric condition
follows from the projection onto ${\cal S}_t$ of one component of Einstein equation and 
results in the elliptic equation \cite{Eardl98}
\be
\label{e:trap_hor}
\left[- {}^2\!\w{\Delta} - 2 \w{L}\cdot {}^2\!\w{D} + A\right] (\beta^\perp-N) = B (\beta^\perp+N) \ \ ,
\ee
where $L_i \equiv K_{kl} s^k q^l_{\ \, i}$, 
$A \equiv  \frac{1}{2} {}^2\!R - {}^2\!\tilde{\w{D}}\cdot \w{L}- \w{L}\cdot \w{L} 
	- 8\pi \w{T}(\hat{\w{\ell}},\hat{\w{k}})$, 
$B \equiv \frac{1}{2} \sigma_{ij}^{(\hat{\w{\ell}})} \sigma^{(\hat{\w{\ell}})ij}
	+ 4\pi \w{T}(\hat{\w{\ell}}, \hat{\w{\ell}})$,
$\w{T}$ is the stress-energy tensor, 
$\hat{\w{\ell}} = \w{n}+ \w{s}$, $\hat{\w{k}} = (\w{n}- \w{s})/2$,
and ${}^2\!\w{D}$, ${}^2\!\w{\Delta}$ and ${}^2\!R$ are respectively the 
covariant derivative, Laplacian and Ricci scalar of $({\cal S}_t,\w{q})$.
The non-negative character of the rhs term in (\ref{e:trap_hor}), together with the lhs elliptic
operator under the FOTH condition (closely related to the stability condition in \cite{AnderMS05}), 
guarantees the positivity of $(\beta^\perp-N)$ in (\ref{e:adapted_coord}). Moreover, 
null-like condition $\beta^\perp=N$ \cite{Cook,JaramGM04} is recovered in the stationary IH limit, 
for which
$\w{\sigma}^{(\hat{\w{\ell}})}=0=\w{T}(\hat{\w{\ell}}, \hat{\w{\ell}})$.
Condition (\ref{e:trap_hor}) provides a relation between combinations $(\beta^\perp-N)$ and $(\beta^\perp+N)$:
given one, the other is fully determined.

\noindent \emph{ii) Gauge conditions for the tangential part of the shift.} Let us express 
the shear tensor along $\w{h}$, $\w{\sigma}^{(\w{h})}$, using the 
coordinate system (\ref{e:adapted_coord}) adapted to ${\cal H}$: 
\bea
2\sigma_{ij}^{(\w{h})} & = & 
 \left(\der{q_{ij}}{t} - \der{}{t}\ln\sqrt{q}\; q_{ij}\right) \nonumber  \\
	& & + \left({}^2\!D_i V_j + {}^2\!D_j V_i - {}^2\!D_k V^k\, q_{ij}\right) 
	\ \ . \label{e:shear_h}
\eea
Imposing as a coordinate choice the vanishing 
of the first parenthesis in the rhs
results in
\be
\label{e:shear_dyn}
{}^2\!D_i V_j + {}^2\!D_j V_i - {}^2\!D_k V^k \, q_{ij} = 2 \sigma_{ij}^{(\w{h})} \ \ ,
\ee
an elliptic equation whose source is determined
by the evolution equation of the shear $\w{\sigma}^{(\w{h})}$ (tidal equation):
\bea
{\cal L}_{\w{h}}\, \sigma^{(\w{h})}_{ij} &=& - {q^k}_i {q^l}_j \ell_m \ell^n {W^m}_{knl}
	- C^2 {q^k}_i {q^l}_j k_m k^n {W^m}_{knl} \nonumber\\
	&-& 8\pi C \left[ {q^\mu}_i {q^\nu}_j T_{\mu\nu} - \frac{1}{2} (q^{\mu\nu}T_{\mu\nu}) q_{ij}\right]
	+ \cdots,
\eea
where $\w{W}$ is the Weyl tensor. Condition (\ref{e:shear_dyn})  
fixes the tangential part of the shift $\w{V}$
up to a linear combination the six conformal symetries in the kernel of the elliptic operator in the lhs.
We determine this conformal symmetry in the evolution by continuity with the conformal Killing 
symmetry prescribed on the initial data.
In the stationary limit, where $\w{h}$ tends to $\w{\ell}$ and $\w{\sigma}^{(\w{\ell})}=0$, the
vanishing of the rhs in Eq. (\ref{e:shear_dyn}) leads to the conformal Killing condition
on $\w{V}$ and, given the rescaling properties of the conformal Killing operator, IH condition
for $\w{V}$ in Refs. \cite{Cook,JaramGM04,GourgJ06} is recovered.

\noindent \emph{iii) Slicing condition.} 
Combined results in Refs. \cite{AshteG05,AnderMS05} show that,
for different choices of 3-foliation $(\Sigma_t)$, 
a given MTS ${\cal S}_t$ on a given initial 3-slice evolves generically into distinct DHs. 
However, all these DHs are 
ultimately expected to approach the event horizon, and therefore there is no preferred candidate 
on the sole basis of the dynamical trapping horizon framework. The choice of inner
BC for $N$, must be adopted on the basis 
of the well-posedness of the elliptic-hyperbolic system and the specific numerical needs. 
In practice, this issue 
must be numerically addressed.
Having said this, Eq. (\ref{e:trap_hor}) suggests an alternative in this context: 
prescribing an inner BC for 
$(\beta^\perp - N)$ determines $(\beta^\perp + N)$ algebraically. Such is the case of the
proposal in \cite{GourgJ06b}, where the choice of that DH locally maximizing 
the area rate of change of
the slice ${\cal S}_t$ 
leads to: $\beta^\perp-N = -\mathrm{const}\cdot\theta^{(\hat{k})}$, with $\mathrm{const}>0$.
Note that if, alternatively, inner conditions are provided for $(\beta^\perp+N)$ [resp. $N$], 
then Eq. (\ref{e:trap_hor}) must be 
solved as an elliptic equation on ${\cal S}_t$ 
for $(\beta^\perp-N)$ [resp. $\beta^\perp$].

\noindent\emph{Inner BCs for the hyperbolic part}. Assessing the freedom 
in prescribing inner BCs for Eqs. (\ref{e:hyperbolic_part}) is a key step in the 
implementation of the fully-constrained evolution scheme. A first analysis 
of the general issues concerning hyperbolicity in Eqs. (\ref{e:hyperbolic_part}), 
has been carried out in Ref.~\cite{CordeIJG07} by writing down the evolution equations
as a first-order system in conservative form, i.e. 
$\partial_t \w{U} +\w{A}^i(\w{U})\partial_i \w{U} = \w{F}[\w{U},...]$,
where the evolving variable vector $\w{U}$ is given by $\w{U}= (\w{h}, \partial_t\w{h}, {\cal D}\w{h})$ 
and matrices $\w{A}^i$ are straightforwardly
determined from Eqs. (\ref{e:hyperbolic_part}). First, it is shown that imposing
the Dirac gauge in (\ref{e:gauges}) indeed guarantees
the real character of the eigenvalues corresponding to matrices $\w{A}^i$, and therefore the
hyperbolicity of the evolution system.
Of particular relevance for the present inner
BC discussion 
is the explicit determination of the (non-vanishing) characteristic speeds associated 
with the vector $\w{s}$ normal to the excision surface ${\cal S}_t$, resulting in 
\cite{CordeIJG07}
\be
\label{e:characteristics}
\begin{array}{ll}
 \lambda^{(\w{s})}_{\pm} = -\beta^\perp \pm N \ & \ \hbox{(each one of multiplicity 6)}. 
\end{array}
\ee
Taking into account the inequality in (\ref{e:adapted_coord}), consequence of the choice of a 
coordinate system adapted to the DH ${\cal H}$  
by enforcing condition (\ref{e:trap_hor}), we conclude the absence of ingoing radiative modes into
the integration domain $\Sigma_t$ at the excision surface. Therefore no inner BC
whatsoever must be prescribed for the hyperbolic part, as a consequence of our choice of 
BCs for the elliptic part. This confirms our initial motivation for using space-like
excision worldtubes in the evolution and shows the key interplay between elliptic and hyperbolic modes
in the coupled fully-constrained evolution system.

\noindent{\em Discussion.} 
In the context of constrained schemes for excised black hole evolutions 
such as Refs. \cite{BonazGGN04,Moncr07},
inner BCs (\ref{e:BC_Psi}) and (\ref{e:trap_hor}), together with the 
essentially free choice of 3-slicing, characterize the inner excision hypersurface ${\cal H}$
as a world-tube sliced by a family $({\cal S}_t)$ of MOTS. 
If, in addition, the 
condition (\ref{e:shear_dyn}) is enforced, 
then IH inner BCs \cite{Cook,JaramGM04,DainJK05,GourgJ06,JaramAL07} 
are recovered in the stationary limit and one of our basic requirements is fulfilled.
Even though the excision world-tube ${\cal H}$ is 
indeed expected to be a DH in realistic contexts, such a character is not actually enforced since the MTS
condition $\theta^{(\w{k})}<0$ is not explicitly imposed. This is not a shortcoming of the approach. 
In fact, an (arbitrary) negative value for $\theta^{(\w{k})}$ could be explicitly 
enforced as a Robin condition 
on $\beta^\perp\cdot \Psi^2$ (cf. Eq. (16) in Ref. \cite{JaramAL07}): together with Eqs. 
(\ref{e:BC_Psi}), (\ref{e:trap_hor}) this would fix $N$, therefore providing an alternative
manner of fixing the slicing. 
However it is known that the future evolution of a DH can cease ``momentarily'' to satisfy MTS and FOTH
conditions, e.g. in the merging of two black holes once the common horizon has shown up.
In this situation, insisting in the prescription of a negative $\theta^{(\w{k})}$ probably
leads to the ill-posedness of the whole coupled elliptic-hyperbolic system. For this reason, 
we rather adopt the methodological choice of only prescribing MOTS as inner BCs.
Regarding a possible FOTH condition failure, and according with the characteristic analysis in \cite{CordeIJG07},
monitoring the sign of $(\beta^\perp - N)$
determines if inner BCs must or must not be provided for the radiative modes. 
This work  
represents an intermediate step 
in the ongoing program \cite{BonazGGN04} addressing
fully-constrained 
excised black hole numerical evolutions.

\noindent This work has been supported by the Marie Curie contract MERG-CT-2006-043501
(J.L.J.), the doctoral fellowship AP2005-2857 from MEC (I.C-C.), 
and grants AYA2004-08067-C03-01 from MEC, 
HF2005-0115 from CNRS/MEC, and 06-2-134423 MATH-GR from ANR.

\end{document}